\documentclass[preprint]{aastex}

\newcommand{\be}{\begin{equation}}
\newcommand{\ee}{\end{equation}}
\newcommand{\crcon}{ {\cal C}_r } 
\newcommand{\mout}{{ m_{\rm out} }} 
\newcommand{\diffuse}{{ {\cal D} }} 
\newcommand{\erf}{{ {\rm Erf} }} 
\newcommand{\ebar}{{ \langle E \rangle }} 
\newcommand{\pbound}{{ {\cal P}_{\rm bound} }} 
\newcommand{\jrms}{{ j_{\rm rms} }} 
\newcommand{\etarms}{{ \eta_{\rm rms} }} 
\newcommand{\pback}{{ \rho_{\rm ret} }} 
\newcommand{\norb}{{ N_{\rm orb} }} 
\newcommand{\eigen}{{ \Lambda }} 
\newcommand{\reduce}{{\Gamma_R}}

\begin{document} 

\title{\bf TURBULENCE IN EXTRASOLAR PLANETARY SYSTEMS \\ 
IMPLIES THAT MEAN MOTION RESONANCES ARE RARE}   


\author{Fred C. Adams\altaffilmark{1,2}, Gregory Laughlin\altaffilmark{3}, 
and Anthony M. Bloch\altaffilmark{1,4} } 

\affil{$^1$Michigan Center for Theoretical Physics \\
Physics Department, University of Michigan, Ann Arbor, MI 48109} 

\affil{$^2$Astronomy Department, University of Michigan, Ann Arbor, MI 48109} 

\affil{$^3$Lick Observatory, University of California, Santa Cruz, CA 95064} 

\affil{$^4$Department of Mathematics, University of Michigan, 
Ann Arbor, MI 48109}

\begin{abstract} 

This paper considers the effects of turbulence on mean motion
resonances in extrasolar planetary systems and predicts that systems
rarely survive in a resonant configuration.  A growing number of
systems are reported to be in resonance, which is thought to arise
from the planet migration process.  If planets are brought together
and moved inward through torques produced by circumstellar disks, then
disk turbulence can act to prevent planets from staying in a resonant
configuration. This paper studies this process through numerical
simulations and via analytic model equations, where both approaches
include stochastic forcing terms due to turbulence.  We explore how
the amplitude and forcing time intervals of the turbulence affect the
maintenance of mean motion resonances. If turbulence is common in
circumstellar disks during the epoch of planet migration, with the
amplitudes indicated by current MHD simulations, then planetary
systems that remain deep in mean motion resonance are predicted to be
rare. More specifically, the fraction of resonant systems that survive
over a typical disk lifetime of $\sim$1 Myr is of order 0.01. If mean
motion resonances are found to be common, their existence would place
tight constraints on the amplitude and duty cycle of turbulent
fluctuations in circumstellar disks. These results can be combined by
expressing the expected fraction of surviving resonant systems in the
approximate form $\pbound \approx C / \norb^{1/2}$, where the
dimensionless parameter $C \sim 10 - 50$ and $\norb$ is the number of
orbits for which turbulence is active.

\end{abstract} 

\keywords{MHD --- planetary systems --- planetary systems: formation ---
planets and satellites: formation --- turbulence}

\section{INTRODUCTION}  

An increasing number of the observed extrasolar planetary systems have
been discovered to contain multiple planets. A growing subset of these
multiple planet systems have period ratios close to the ratios of
small integers and hence are candidates for being in mean motion
resonance (e.g., Mayor et al. 2001, Marcy et al. 2002, Butler et al.
2006).  In true resonant configurations, the orbital frequencies not
only display integer ratios, but, in addition, the relevant resonant
arguments (angular variables composed of the osculating orbital
elements) display constrained oscillatory time evolution (for further
discussion, see Murray \& Dermott 1999; hereafter MD99; see also Peale
1976).  As a result, these resonant configurations represent rather
special dynamical states of the planetary system and their existence
(if/when observed) places interesting constraints on the formation and
long term evolution of these systems. More specifically, as we show in
this paper, planetary systems are easily knocked out of mean motion
resonance. One source of perturbations acting on the planets is
turbulent fluctuations in the circumstellar disks that originally
formed the planets.  Since these disks are also necessary ingredients
in the current picture of planetary migration (e.g., Papaloizou \&
Terquem 2006), and since coupled migration seems necessary to put
planets into mean motion resonance (e.g., Lee \& Peale 2002; Lee 2004;
Moorhead \& Adams 2005; Beaug{\'e} et al. 2006), the effects of disk
turbulence on the resonances is important to understand.

This paper analyzes the effects of turbulence on planets in or near
mean motion resonance using both a semi-analytic treatment (\S 2) and
direct numerical integrations (\S 3). The semi-analytic approach makes
extreme approximations in order to obtain tractable equations, but it
allows for an explicit determination of how the long term behavior
depends on the basic physical variables (e.g., planet masses,
semimajor axes, type of resonance, turbulent forcing amplitude, etc.).
In contrast, numerical integrations can be carried out to high
accuracy including all 18 dynamical variables of the problem (for
3-body systems).  Although the simulations demonstrate what the long
term behavior will be, they don't automatically specify the
relationships between the aforementioned physical variables.
Fortunately, both approaches give consistent results for the
intermediate time scales of interest. In particular, we find that the
expectation value of the effective energy of the resonance --
considered as a nonlinear oscillator -- increases linearly with time.
Similarly, the probability that systems stay in resonance decreases
with time. For the stochastic pendulum model, the fraction of systems
in resonance decreases as the square root of time for the simplest
case in which the systems freely random walk in and out of a bound
state; the decay is exponential if systems cannot return to resonance
after leaving.  The full numerical integrations give results that are
intermediate between these extremes.  For the model equations, the
leading coefficients in the time-dependent relationships are given by
the amplitude of the fluctuations, the time intervals of their
forcing, and the natural oscillation frequency of the libration angle
of the original resonance. This paper thus provides a relatively
simple description of the effects of turbulent fluctuations on
planetary systems in or near mean motion resonance.

\section{SEMI-ANALYTIC MODEL EQUATIONS} 

\subsection{Pendulum Model for the Resonance} 

For the sake of definiteness, we consider the case in which a larger
planet on an external orbit perturbs a smaller planet on an interior
orbit. The larger body is then assumed to have orbital elements that
do not change with time (or at least vary much less than those of the
smaller inner planet).  We are thus making the most extreme
approximation and keeping only the leading order terms in order to
obtain an analytic description.  Following MD99, the equation of
motion for the libration angle $\varphi$ for two planets in resonance
reduces to that of a pendulum, i.e.,
\be 
{d^2 \varphi \over dt^2} + \omega_0^2 \sin \varphi = 0 \, , 
\label{eq:pendone} 
\ee 
where the natural oscillation frequency $\omega_0$ is given by 
\be 
\omega_0^2 = - 3 j_2^2 \crcon \Omega e^{|j_4|} \, .
\label{eq:omega} 
\ee
Here, $\Omega$ and $e$ are the mean motion and eccentricity of the
inner planet. The integers $j_2$ and $j_4$ depend on the type of
resonance being considered. The parameter $\crcon$, which is taken 
to be constant, is given by 
\be
\crcon = \mu \Omega \, \alpha f_d(\alpha) \, , 
\ee
where $\mu$ = $\mout/M_\ast$ and where $\mout$ is the mass of the
outer (perturbing) planet.  The quantity $\alpha f_d (\alpha)$ results
from the expansion of the disturbing function (MD99), where the
parameter $\alpha$ is the ratio of the semimajor axes of the two
planets, i.e., $\alpha \equiv a_1/a_2$. This approximation neglects
terms of order ${\cal O}(\mu)$. Finally, we note that more complicated
analytic models for mean motion resonances can be derived (e.g.,
Holman \& Murray 1996; Quillen 2006), but they are qualitatively
similar to the pendulum equation considered here. 

For much of this analysis, we are interested in the 2:1 mean motion
resonance since observed extrasolar planetary systems are (sometimes)
found in or near such a configuration, and since these resonances are
generally the strongest.  For this case, the integers $j_2 = -1$ and
$j_4 = -1$.  For the 2:1 resonance, $\alpha f_d(\alpha) \approx -3/4$,
where $\alpha$ is the ratio of semimajor axes of the two planets. With
these specifications, the natural oscillation frequency $\omega_0$ of
the libration angle is given approximately by
\be
\omega_0^2 \approx {9 \over 4} \mu \Omega^2 \, e \,  . 
\label{eq:omegadef} 
\ee 
In typical cases, where planet masses are 1000 times smaller than 
the stellar masses, and the eccentricity $e \sim 0.1$, we find 
that $\omega_0 / \Omega \sim 10^{-2}$. As a result, the period of 
oscillation for the libration angle is typically $\sim100$ orbits. 

For completeness, we note that for first order resonances, the
oscillation frequency $\omega_0^2$ often has additional contributing
terms (MD99).  The order of the additional terms differs from that of
equation (\ref{eq:omega}) by a factor of order ${\cal O} (\mu / e^3)$,
where $e$ is the eccentricity of the inner planet. Under some
circumstances of interest, $\mu \sim 10^{-3}$ and $e \sim 0.1$, so
this correction can be of order unity. We are thus considering only
the simplest form of the pendulum equation in this analysis, in order
to gain a general understanding of the effects of turbulence on
resonant systems.  In principle, however, care must be taken to
include all of the relevant terms.

\subsection{Turbulent Perturbations} 

Given the pendulum model for the resonance (from MD99), the next step
is to include the effects of additional perturbations produced by
turbulence in the circumstellar disk material that remains after the
planets have been formed. Since this disk is generally thought to
induce planetary migration, some disk material is likely to be
present, for a typical time scale of 1 -- 10 Myr. Under many
circumstances, these disks are expected to be turbulent (Balbus \&
Hawley 1991) due to the magneto-rotational instability (MRI). Although
MRI can be shut down if the ionization rate is not high enough (Gammie
1996), star forming regions can experience enhanced cosmic ray fluxes,
and hence enhanced ionization rates, due to cosmic rays from
supernovae becoming trapped in the magnetic fields of molecular clouds
(Fatuzzo et al.  2006). In this work, we consider turbulent
fluctuations to be present and consider our previous model (Laughlin
et al. 2004; hereafter LSA) to provide the parameters of the
perturbations. Note that the LSA model describes turbulence is disks 
with no gaps, applicable to low-mass embedded planets, whereas this 
paper primarily considers giant planets that will clear gaps. As a 
result, we must modify the LSA formalism to include this effect 
(see below). 

In principle, the turbulent fluctuations in the disk can affect either
planet, i.e., the outer perturbing planet (in this analysis) or the
inner perturbed planet. If the outer planet is subject to turbulent
fluctuations, they have the effect of shaking the effective base of
the pendulum. As a result, they modify the effective potential
energy of the pendulum, and the stochastic term enters into the
equation of motion as follows:
\be
{d^2 \varphi \over dt^2} + [\omega_0^2 + \xi(t)] 
\sin \varphi = 0 \, , 
\label{eq:outerpend} 
\ee
where $\xi(t)$ is the stochastic forcing term. This is the form 
expected when the circumstellar disk surrounds both planets and 
primarily influences the outer one. In contrast, if the disk 
(the torques produced by turbulence in the disk) primarily 
influences the inner planet, then the fluctuations provide an 
additional acceleration term and the equation of motion takes 
the form 
\be 
{d^2 \varphi \over dt^2} + \omega_0^2 \sin \varphi = 
- {\widetilde \xi(t)} \, . 
\label{eq:innerpend} 
\ee
For the sake of simplicity, we will primarily consider the first 
case (eq. [\ref{eq:outerpend}]). Notice also that both types of 
fluctuations can be included. 

The inclusion of turbulent fluctuations thus provides a stochastic
forming term in the pendulum equation of motion (\ref{eq:pendone}).
The forcing term $\xi(t)$ can be considered as a series of ``kicks''
produced by short-term forcing due to turbulence. To completely
specify this forcing effect, one needs to determine the spectrum of
fluctuation amplitudes, the timing of the fluctuations, and any
possible correlations in the fluctuation timing. This latter property
determines the so-called ``color'' of the noise (fluctuations) and can
have (perhaps) surprising effects on the long-term evolution of the
stochastic pendulum and hence the resonance angles of the planetary
system (Mallick \& Marcq 2004, hereafter MM04).  

The formulation for turbulence developed in LSA shows that the
turbulent fluctuations are reloaded into the disk on roughly an
orbital time scale. In order to consider the turbulent torques to be
fully independent, we must use the longest relevant orbital period,
which corresponds to that of the outer disk edge (in the calculations
of LSA).  This period at the outer disk edge is about twice that of
the planet being forced. Since the outer (perturbing) planet is being
forced by the turbulence in the present context, and since the outer
planet is in a 2:1 mean motion resonance with the inner planet, this
turbulent forcing period is approximately four times the period of the
inner planet.

Next we need to account for the fluctuations in the equation of
motion.  To the same order of approximation used to derive the
pendulum equation for the resonance, the time rate of change of the
mean motion is given approximately by
\be
{d \Omega \over dt} \approx - {1 \over 3} \Omega 
\left( { 1 \over J} {d J \over dt} \right) \, ,  
\ee 
where $J$ is the angular momentum of the inner planetary orbit. 
Here we consider the turbulent fluctuations to provide discrete
changes in the angular momentum on time scales comparable to the
orbital period $P_D$ at the disk edge, where (as discussed above) we
expect $P_D \approx 4 P_{orb} = 8 \pi/\Omega$. Thus, the time scale
$\Delta t$ for independent stochastic perturbations to act is given by
$\Delta t \approx 8 \pi/\Omega$. Since $d^2 \varphi/dt^2 \sim d\Omega
/dt$ (see equation [8.39] from the derivation of MD99), the forcing
term $\xi_k \propto \Omega ({\dot J}/J) \sim \Omega [(\Delta J)_k/J]
\delta ([t] - \Delta t)$. As a result, we can write the stochastic 
differential equation in the form 
\be
{d^2 \varphi \over dt^2} + \left[ \omega_0^2 - 
{1 \over 3} \Omega \left( {\Delta J \over J} \right)_{k} 
\delta \left([t] - \Delta t \right) \right] \sin \varphi = 0 \, .
\ee 
In this equation, $[(\Delta J)/J]_k$ represents the fractional change
in the angular momentum of the planet for a given realization of the
turbulence. The subscript `$k$' indicates that this increment of
angular momentum is delivered at discrete time intervals, which can be
counted and labeled with an integer. To account for this timing, 
the Dirac delta function is periodic with period $\Delta t$, so that  
the quantity $[t]$ is the time measured mod-$\Delta t$. 

The amplitude $[(\Delta J)/J]_k$ is calculated in LSA for disks
without gaps (see also Nelson \& Papaloizou 2003, hereafter NP03, and
the discussion below). The resulting torques produce angular momentum
perturbations with zero mean and well-defined RMS amplitudes of order
$[(\Delta J)/J]_{\rm rms} = \jrms \sim 0.004$. Other three dimensional
MHD simulations produce fluctuations with comparable amplitudes (e.g.,
Nelson 2005).

For planets that clear gaps in the disk, as considered here, we must
include a reduction factor $\reduce$ to account for the absence of the
full complement of disk material near the planet. In practice, we find
that $\reduce \approx 0.1$, where this value can be estimated in
multiple ways. First, we note that numerical simulations of gap
formation in turbulent disks (NP03) show that the gaps are not
completely cleared out; instead, the surface density at the gap
minimum is reduced by a factor of $\sim20$ from the unperturbed level
(see Fig. 1 of NP03). Thus we expect the reduction factor to lie in
the range $0.05 \le \reduce \le 1$. Next we note that material from
both the outer part of the gap (where the reduction in surface density
is less severe) and outside the gap (where there is no reduction)
provide some contribution to the torques. Indeed, the dominant
contribution of torques is expected to come from intermediate length
scales (see Johnson et al. 2006), specifically scales larger than the
disk scale height, but not too much larger (given that gravitational
forces decrease with distance). To model this effect, we start with
the formalism of LSA and remove disk material corresponding to a
quadratic gap profile with zero surface density at the center (e.g.,
Goldreich \& Sari 2003) and then calculate the reduction factor
$\reduce$ as a function of gap width.  Let $w_g$ denote the half-gap
width, i.e., the location on either side of the planet where the
quadratic gap profile joins onto the unperturbed surface density
distribution. The planet is assumed have semi-major axis $a_p$, which
also defines the location of the gap center, where the surface density
is taken to vanish.  Then we find that $\reduce \approx$ 0.2, 0.1, and
0.05 for total gap width $2 w_g / a_p$ = 1, 1.5, and 2, respectively.
Although the exact value of the reduction factor will depend on the
details of the gap shape, we use $\reduce$ = 0.1 as a representative
value for this work. We also note that these systems have two planets,
so that each planet can clear disk material, but each planet can also
experience torques; we assume here that these two competing effects
cancel.

For completeness, we note that the surface density structure in the
local neighborhood of a massive planet can be strongly perturbed, with
well-defined wakes and even a circumplanetary disk within the gap
(e.g., see Fig. 10 of NP03). Although global transport seems to be
unaffected by these distortions, their effects on stochastic torques
remain unclear and should be considered in future work.

Since the forcing amplitude $[(\Delta J)/J]_k$ plays an important
role, it is useful to have an heuristic understanding of the expected
value. For a circumstellar disk with surface density $\Sigma$, the
torque exerted on a planet can be written in terms of the benchmark
value $T_D = 2 \pi G \Sigma r m_{\rm p}$.  The torque due to
turbulence is expected to be a fraction $f_T$ of this physical scale,
where numerical simulations suggest that $f_T \approx 0.05$ (e.g.,
Johnson et al. 2006; Nelson 2005).  This torque acts over a time
interval to produce a change in angular momentum. As argued above,
this calculation uses a time interval of four orbital periods, long
enough so that the turbulent torques can be taken to be independent.
The torque thus acts on a timescale $4 P_{orb} = 8 \pi (r^3 / G
M_\ast)^{1/2}$ and the net change in angular momentum $(\Delta J) = 4
P_{orb} f_T T_D$. Next, we include the reduction factor $\reduce$ due
to the disk gap (where we expect $\reduce \sim 0.1$).  Finally, the
orbital angular momentum $J$ of the planet is given by $J = m_{\rm p}
[G M_\ast r (1-e^2)]^{1/2}$. We can absorb the eccentricity dependence
into the definition of $f_T$.  Putting these results together, we can
write the fractional change in angular momentum over one time interval
($4 P_{orb}$) in the form
\be
\left( { \Delta J \over J} \right)_k = f_T \reduce \, 
{16 \pi^2 \Sigma r^{2} \over M_\ast} \, \approx \, 10^{-4} \, 
\left( {\Sigma \over 1000 {\rm g} \, {\rm cm}^{-2} } \right) \, , 
\label{eq:bench} 
\ee 
where the second equality is scaled for a 1 AU orbit and a typical
surface density at that radius. Note that we expect the surface
density profile to have a nearly power-law form $\Sigma \sim r^{-p}$,
where $p \sim 3/2$, so the quantity $\Sigma r^2$ increases slowly 
with radius. Since giant planets form at larger radial 
location and migrate inwards, the torques will be somewhat larger 
during the earlier phases of migration. 

\subsection{Time Evolution of the Stochastic Pendulum} 

With the mean motion resonance modeled by a pendulum equation, and
with the characteristics of the turbulent fluctuations specified, we
now find the time evolution of the system.  If we define a
dimensionless time variable 
\be
\tau \equiv \omega_0 t \, , 
\ee
where $\omega_0$ is the oscillation frequency of the libration angle,
and is defined by equation (\ref{eq:omegadef}), the equation of
motion takes the form 
\be 
{d^2 \varphi \over d\tau^2} + \left[ 1 + \eta_k 
\delta \left( [\tau] - \Delta \tau \right) \right] 
\sin \varphi = 0 \, , 
\ee 
where $\delta (x)$ is the Dirac delta function. In this formulation,
$\Delta \tau$ is the time interval between stochastic kicks, the
dimensionless time variable is measured mod-$\Delta \tau$, and the
amplitude of the kicks is given by
\be
\eta_k = - {1 \over 3} {\Omega \over \omega_0} 
\left( {\Delta J \over J} \right)_{k} \, = 
- {2 \over 9 \sqrt{\mu e}} \left( {\Delta J \over J} \right)_{k} \, . 
\label{eq:etadef} 
\ee
The quantity $\eta_k$ is thus a stochastic variable that has a
distribution of values inherited from the distribution of values of
$[(\Delta J)/J]_k$, which in turn is determined by the properties of
the turbulence. The kicks in angular momentum can be either positive
or negative, with no expected asymmetry, so that $\langle \eta_k
\rangle$ = 0. We can thus characterize the amplitude by the RMS
(root-mean-square) of the variable.  Using the amplitudes expected
from numerical MHD simulations (\S 2.2), this amplitude is expected 
to lie in the range 
\be
\etarms = \langle \eta_k^2 \rangle^{1/2} \approx 0.002 - 0.009 \, , 
\label{eq:etarms} 
\ee 
where we have used typical values of the mass ratio $\mu = 10^{-3}$
and the eccentricity $e$ = 0.1 to evaluate the result. The typical
expected value of the RMS fluctuation amplitude is thus $\etarms \sim
0.005$. These acceleration kicks occur at typical time intervals
$\Delta t \approx 8 \pi / \Omega$, which sets the corresponding
dimensionless time interval to be $\Delta \tau \approx 8 \pi
\omega_0/\Omega$.  In general, both the amplitudes and the time
intervals between forcing kicks will vary from cycle to cycle
according to their (respective) distributions. However, we can scale
out the time variations (Appendix A; Adams \& Bloch 2008) and
characterize the turbulence with a well-defined (single-valued)
forcing time. As a result, we present the remainder of this analysis
using a single forcing time interval and focus on the effects of the
distribution of forcing amplitudes $\eta_k$.

\begin{figure} 
\figurenum{1} 
{\centerline{\epsscale{0.90} \plotone{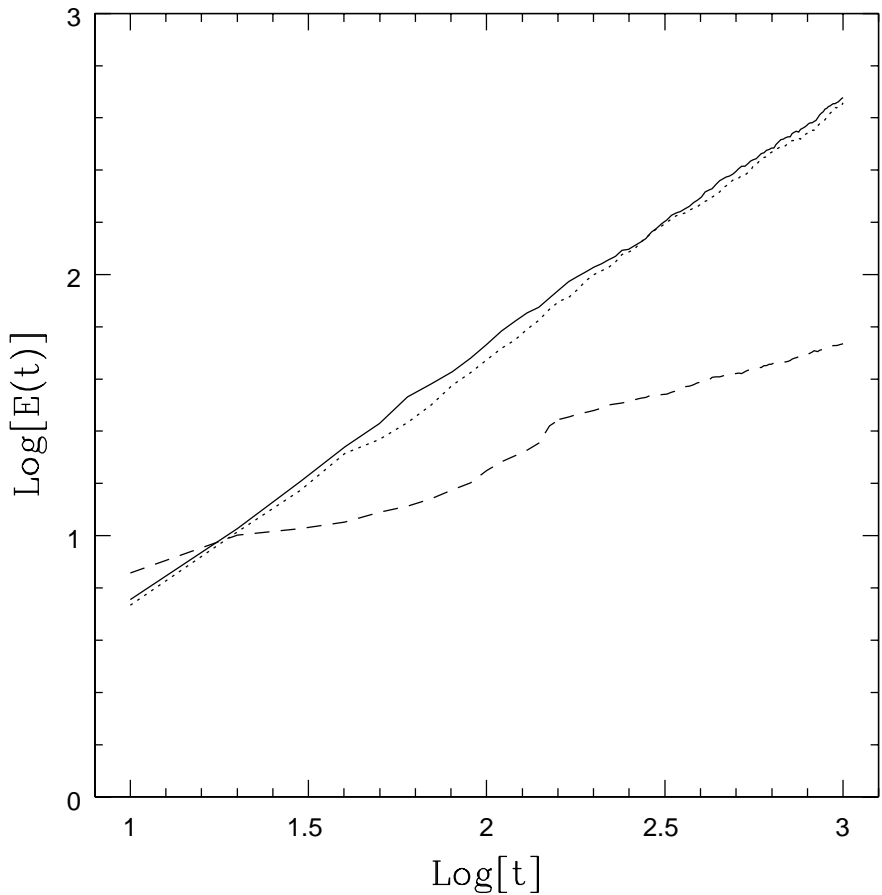} } } 
\figcaption{Mean time evolution of the energy of a stochastic pendulum
with noise terms of various ``colors''. For each curve/line, $10^4$ 
realizations of the time evolution of the oscillator have been
averaged together. The curves show the results for white noise
(solid), pink noise (dotted), and brown noise (dashed). The first two
types of noise lead to energy evolution that is linear in time, $\ebar
\sim t$, whereas brown noise leads to energy that increases $\ebar
\sim \sqrt{t}$. } 
\label{fig:enertime}
\end{figure} 

To analyze the behavior of the stochastic pendulum, as defined above, 
it is useful to work in terms of phase space variables. We thus rewrite 
the equation of motion into two parts, i.e., 
\be 
{d \varphi \over d \tau} = V \, , \qquad {\rm and} \qquad 
{d V \over d \tau} = - \left[ 1 + \eta \right] \sin \varphi \, . 
\ee 
We then define the probability distribution function for the phase
space variables, $P(\varphi, V; t)$, which obeys the Fokker-Planck
equation (MM04; Binney \& Tremaine 1987) 
\be 
{\partial P \over \partial \tau} + V {\partial P \over \partial \varphi} 
- \sin \varphi {\partial P \over \partial V} = {\diffuse \over 2}  
\sin^2 \varphi {\partial^2 P \over \partial V^2 } \, , 
\ee
where the phase space diffusion constant $\diffuse$ is set by the
amplitude of the fluctuation spectrum.  Specifically, we can write 
the diffusion constant in terms of previously defined variables, 
\be
\diffuse = {\langle \eta_k^2 \rangle \over \Delta \tau} \approx 
{\langle \eta_k^2 \rangle \Omega \over 8 \pi \omega_0} \, . 
\ee 
Next we argue that the libration angle $\varphi$ varies rapidly
compared to the velocity $V$. This claim can be verified by noting
that ${d V /d\tau} \sim \eta_k$ so that $V \sim \tau^{1/2}$ (since the
random variable $\eta_k$ implies a random walk growth). However, the
libration angle obeys the equation $d \varphi / d \tau = V$, which in
turn implies that $\varphi \sim \tau^{3/2}$ (see MM04 for further
discussion).  In the long term, the libration angle thus varies faster
than $V$ and we can average the Fokker-Planck equation over the angle
$\varphi$ to obtain the time evolution equation for the averaged
probability distribution function $p(V; \tau)$, i.e., 
\be
{\partial p \over \partial \tau} = \langle \sin^2 \varphi \rangle 
{\diffuse \over 2} {\partial^2 p \over \partial V^2} \, . 
\ee 
This equation has the well-known solution 
\be
p(V; \tau) = {1 \over (\pi D \tau)^{1/2} } 
\exp \left[ - {V^2 \over D \tau} \right] \, , 
\ee
where we have defined an effective diffusion constant $D = 2 \diffuse
\langle \sin^2 \varphi \rangle$.  In the long time limit, $\varphi
\sim \tau^{3/2}$, and the angle fully samples the range $[0, 2\pi]$,
so that $\langle \sin^2 \varphi \rangle = 1/2$ and hence $\diffuse =
D$.  Notice also that the velocity $V$ grows as $V \sim \tau^{1/2}$ in
the long time limit. Since the energy of the pendulum grows large at
long times, the kinetic term dominates the potential energy term, and
the energy $E \approx V^2/2$.  With this substitution, the probability
distribution function for the energy takes the simple form 
\be
P (E; \tau) = \left( {2 \over \pi D \tau} \right)^{1/2} E^{-1/2} 
\exp \left[ - {2E \over D \tau} \right] \, . 
\label{eq:edistrib} 
\ee
This result implies that the expectation value of the 
energy grows linearly with time in the long time limit, i.e., 
\be
\langle E \rangle = {D \over 4} \tau \, . 
\label{eq:evst} 
\ee 
This linear time dependence can be understood in terms of a random
walk of velocity, as shown in Appendix B.

This time behavior is seen in numerical simulations of the stochastic
pendulum equation. Figure \ref{fig:enertime} shows the time
development of the mean energy, averaged over $10^4$ realizations, for
three different types of fluctuations. The solid curve shows white
noise, where the fluctuation amplitude and the time interval $\Delta
t$ between kicks are uniformly distributed about well-defined mean
values. The dotted curve shows the case of ``pink'' noise, in which
the distribution of time intervals is taken to have a decaying
exponential form; this distribution leads to more time intervals near
the lower end of the spectrum. As long as the mean time interval and
mean fluctuation amplitude remain the same, and as long as both are
uncorrelated with previous samples, the linear time dependence
predicted by equation (\ref{eq:evst}) holds. For completeness, Figure
\ref{fig:enertime} also shows the case of ``brown'' noise, in which
the time intervals undergo a random walk (akin to brownian motion) so
that a correlation of time intervals is present; in this case, the
time development of the energy is slower with $\langle E \rangle
\propto t^{1/2}$ (see also MM04). Thus, the long term evolution of an
ensemble of stochastic pendulums obeys the analytic expectations
derived above, provided that (1) the fluctuations are independent, 
(2) the long-time limit has been reached, and (3) the system can 
freely random walk back into a resonant (bound) state (from a 
higher energy, unbound state). 

\begin{figure} 
\figurenum{2} 
{\centerline{\epsscale{0.90} \plotone{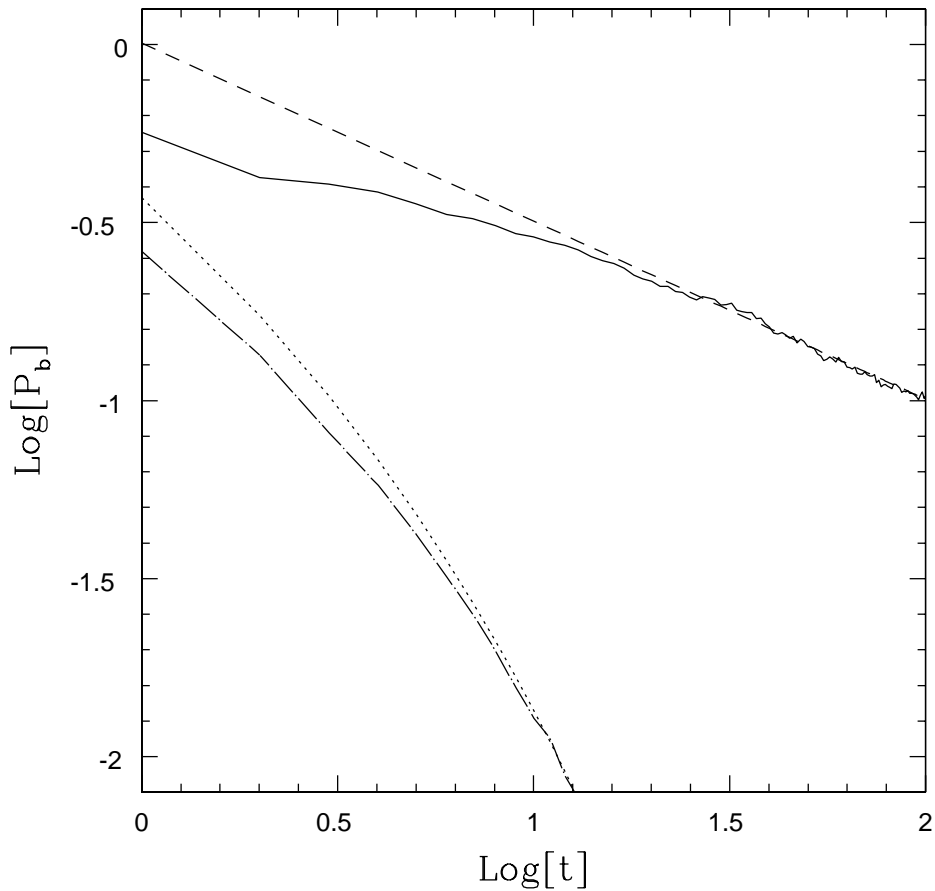} } } 
\figcaption{Time evolution of the probability for the planetary system
to stay in a mean motion resonance as a function of time. The solid 
curve shows the result of $10^4$ realizations of the stochastic 
pendulum equation; in this set of numerical experiments, the system 
is allowed to come in and out of resonance freely. The dashed curve 
shows the expected asymptotic behavior of the system, i.e., a survival 
probability of the form $\pbound \propto t^{-1/2}$. The dot-dashed curve 
shows the survival probability for when a one-way barrier is imposed 
so that systems that leave resonance (the pendulum becomes unbound) 
are not allowed to re-enter the bound configuration. The dotted curve 
shows the analytic approximation to this probability evolution derived 
in the text. }  
\label{fig:pdecay}
\end{figure}  

\subsection{Upper Limit on the Fraction of Bound States} 

Given the time evolution of the distribution function, we now consider
the fraction of low energy states as a function of time.  Here, states
of the system with sufficiently low energy are bound and the pendulum
oscillates --- instead of circulates --- so that the system is in
resonance. We are thus finding an estimate of the fraction of systems
that remain in resonance as a function of time, where systems leave
resonance due to exposure to turbulent forcing. Note that our
treatment thus far provides only an upper limit to the fraction of
bound states because we allow the system to freely random walk both in
and out of resonance.  The fraction of bound states derived in this
section thus represents the largest possible value.  In the following
section, we find a more accurate estimate by accounting for the
reduced probability of systems returning to resonance after leaving.

As formulated here, the energy $E$ is the kinetic energy, so that the
criterion for the resonance to be undone (for the libration angle to
circulate) is given (approximately) by $E > 1$. Note that the system
must have $E > 2$ to be freely circulating, but systems with energies
in the range $1 < E < 2$ have such large libration angles as to not be
in a well-defined resonant state.  As encapsulated in the probability
distribution function for the energy, the system always has some
probability of remaining in resonance, even as the mean energy
$\langle E \rangle$ grows ever larger. This probability is a
decreasing function of time and is given by
\be 
\pbound = 
\left( {2 \over \pi D \tau} \right)^{1/2} \int_0^1 {dE \over \sqrt{E} } 
\exp \left[ - {2E \over D \tau} \right] \, = {2 \over \sqrt{\pi}} 
\int_0^{z_0} {\rm e}^{-z^2} dz \, = \erf (z_0) \, , 
\label{eq:pbint} 
\ee
where $z_0 = (2/D \tau)^{1/2}$. Here $\erf (z)$ is the error function
(e.g., Abramowitz \& Stegun 1970), which can be expanded in the limit 
of small $z$ as follows:
\be
\erf (z) = {2 \over \sqrt{\pi}} \left( z - {z^3 \over 3} + {z^5 \over 10} 
- {z^7 \over 42} + \dots \right) \, ,
\ee
where small $z$ values correspond to late times.  To leading order, the 
probability that the planetary system remains in resonance is thus
given by the relation
\be
\pbound \approx \left( {8 \over \pi D \tau} \right)^{1/2} = 
\left( {2 \over \pi \langle E \rangle} \right)^{1/2} \, = 
{8 \over \etarms \bigl[ \Omega t \bigr]^{1/2} } \, =  
{4 \sqrt{2/\pi} \over \etarms \sqrt{\norb} } \, , 
\label{eq:probtime} 
\ee
where this expression is valid only for sufficiently late times when
$\tau > 2/D$.  For planetary systems, the natural clock is set by the
orbit time, so in the final equality we measure time using the number
of orbits $\norb \equiv \Omega t/(2\pi)$. If we then insert the
expected fluctuation amplitude $\etarms \approx 0.005$ (eq.
[\ref{eq:etarms}]), the above expression simplifies to the form
$\pbound \approx 640 / \norb^{1/2}$, where $\norb$ includes only the
orbits (time) for which turbulence is active.  For typical systems,
where the number of orbits $\norb = 10^6 - 10^8$, this expression
implies that the probability of remaining in resonance lies in the
range $\pbound \approx 0.06 - 0.60$.  Keep in mind, however, that this
derivation assumes that planetary systems can freely random walk both
in and out of resonance, so that equation (\ref{eq:probtime}) represents 
an upper limit to the fraction of surviving resonant systems. But even 
this upper limit demonstrates that turbulence significantly reduces
the expected number of resonant systems. In the more realistic case,
where systems can random walk out resonance more easily than they can
random walk back into resonance (\S 2.5 and \S 3), the fraction of
surviving systems will be much lower. 

Equation (\ref{eq:probtime}) shows that the probability of a planetary
system remaining in resonance should decrease as the square root of
time. This dependence (found analytically) is observed in numerical
simulations using the stochastic pendulum equation.  Figure
\ref{fig:pdecay} shows how the probability of staying in resonance
decreases with time for an ensemble of stochastically driven
oscillators. The solid curves show the percentage of systems
remaining in resonance as a function of time, using a sample of $10^4$
systems. The smooth dashed and dotted curves show the analytic
predictions for the time dependence derived above.

This estimate of the probability for remaining bound is approximate in
several respects. First, we have not included the potential energy
term $(\omega_0^2 \cos \varphi)$ in the calculation of the probability
distribution $P(E; \tau)$. This approximation leads to an uncertainty
of $\sim \sqrt{2}$ in the estimates presented here. More importantly,
we again stress that the analysis presented thus far represents an
upper limit on the fraction of systems that can remain in resonance.
In the stochastic pendulum formalism, we are implicitly assuming that
the system can random walk both in and out of a bound configuration,
i.e., that the planetary system can random walk in and out of
resonance. The model equation developed here only has one degree of
freedom ($\varphi$), so this behavior is natural. In actual planetary
systems, however, other physical variables are relevant and must have
the proper values to allow the planets to be in resonance. This
difficulty implies that once a planetary systems leaves resonance, it
will be unlikely to random walk back into resonance.  This behavior,
in turn, implies that the fraction of planetary systems remaining in
resonance will fall below the fiducial law derived here. 
We consider this complication in the following sections. 

\subsection{Evolution with a One-way Barrier} 

The discussion presented thus far assumes that the stochastic
oscillator can diffuse in and out of a bound state, i.e., into
resonance as well as out of resonance. For the simple model case,
which has only one variable, this behavior is sensible. For real
planetary systems, however, the conditions of resonance are more
complicated than the simple pendulum model. In particular, for highly
interactive systems (e.g., the observed 2:1 resonance in GJ876; see \S
3), many orbital elements of both planets must have the proper values
for the planets to be in a mean motion resonance.  In such systems,
while it remains possible for fluctuations to compromise the
resonance, it will be unlikely for the system to random walk back into
resonance.  When a real planetary system is knocked out of libration
and into circulation, then the chances of returning are reduced
because the chances for close encounters that drastically change the
orbital parameters are significantly increased.  The fraction of
systems that remains bound will thus decrease with time more sharply
than suggested by the considerations of the previous section.  As a
result, the model derived above, where $\pbound \propto \tau^{-1/2}$,
represents an {\sl upper limit} on the fraction of systems that remain
in resonance. In this section, we explore a more complicated model for
the probability evolution that includes the one-way nature of this
process. We also consider the intermediate case of a partially
transmitting barrier, where systems can return to resonance after
leaving, but with reduced probability.

\subsubsection{Heuristic Model Equation} 

We now consider a simple generalization of the time evolution of the
probability. At any given time, the fraction of systems that would
remain bound in the absence of the one-way barrier is given by
$\pbound$ from equation (\ref{eq:probtime}).  We then assume that some
fraction of these systems are ``lost'' at a well-defined rate, so that
the time evolution of the fraction of bound states (in the absence of
the diffusion term) would be an exponential decay. This ansatz thus
assumes that the remaining systems are distributed across the possible
bound energy values, and that each system has some probability of
leaving its bound state per unit time. With this assumption, the
Fokker-Planck equation for the averaged probability distribution
function $p(V; \tau)$ now takes the modified (heuristic) form 
\be
{\partial p \over \partial \tau} = {D \over 4} 
{\partial^2 p \over \partial V^2} - \gamma 
\left( {8 \over \pi D \tau} \right)^{1/2} p \, ,
\label{eq:heuristic} 
\ee 
where we assume that the bound fraction has the asymptotic form
derived above and where the parameter $\gamma$ encapsulates the
details of this ``decay process''. Note that a similar form arises
(Caughey 1963) when dissipation is included in the Fokker-Planck
equation (these connections should be explored further). Equation  
(\ref{eq:heuristic}) can be solved to obtain the result:
\be
p(V; \tau) = {1 \over (\pi D \tau)^{1/2} } 
\exp \left[ - {V^2 \over D \tau} \right] \, \exp \left[ - 
\gamma \left( {32 \tau \over \pi D} \right)^{1/2} \right] \, . 
\ee
With this complication, the fraction of systems that remain bound 
as a function of time now takes the form 
\be
\pbound \approx \left( {8 \over \pi D \tau} \right)^{1/2} \,
\exp \left[ - \gamma \left( {32 \tau \over \pi D} 
\right)^{1/2} \right] \, . 
\label{eq:heursolve} 
\ee 
Although heuristic, this model captures the basic time behavior for
stochastic pendulums with a one-way barrier, as shown in Figure
\ref{fig:pdecay}.

\subsubsection{Solutions from Separation of Variables} 

For the simplest case of a constant diffusion constant, the solution
for a one-way barrier can be found using separation of variables. In
this case, the solution to the diffusion equation is assumed to have
the general form $p(t,V) = T(t) g(V)$; the diffusion equation then can
be written as 
\be
{ 1 \over T} {dT \over dt} {4 \over D} = 
{1 \over g} {d^2 g \over dV^2} = - \eigen^2 \, , 
\ee 
where $\eigen^2$ is the separation constant. 

The solutions $g(V)$  have the form 
\be 
g_n (V) = A_n \cos \eigen_n V \, , 
\ee 
where the subscript denotes one of a series of solutions that satisfy
the boundary condition. We we invoke the boundary condition that
$g(V=\sqrt{2})=0$, which implies that the distribution function
vanishes when the kinetic energy reaches $E = V^2/2 = 1$. This
condition thus specifies the eigenvalues $\eigen_n$, i.e.,
\be
\eigen_n = {\pi \over \sqrt{2} } (n + 1/2) \, , 
\label{eq:eigenvalue} 
\ee
where $n$ is an integer. The full solution thus takes the form 
\be 
p(t,V) = \sum_{n=0}^\infty \, A_n \, \cos (\eigen_n V) \, 
\exp \left[ - {1 \over 4} D \eigen_n^2 t \right] \, . 
\label{eq:sepsolve} 
\ee 
For the initial condition $p(t=0,V)$ = $\delta(V)$, where $\delta(x)$
is the Dirac delta function, the constants are specified so that
\be 
A_n = 1 / \sqrt{2} \, . 
\ee
The fraction of bound states at any time is then given by 
\be
\pbound = 2 \int_0^{\sqrt{2}} p(t,V) dV =  
\sum_{n=0}^\infty {2 (-1)^n \over \pi (n + 1/2)} \, 
\exp \left[ - {D \over 8} \pi^2 (n + 1/2)^2 t \right] \, . 
\ee
This solution, though approximate, indicates that the long term
evolution of the ensemble of resonant states follows an exponential
decay. For the simple solution of equation (\ref{eq:sepsolve}), each
``mode'' decays exponentially, albeit with a differing decay rate. In
the long run, however, only the lowest order mode survives, and the
time evolution becomes purely exponential.

\subsubsection{Further Complications} 

The evolution of the actual stochastic pendulum differs from that
described by a simple diffusion equation in several respects. Most
importantly, the diffusion constant $D$ is not really a constant.  In
the formulation developed here, $D \propto \sin^2 \varphi$, and we
have taken the average value $\langle \sin^2 \varphi \rangle$ = 1/2.
However, this limit only applies in the long time limit when most of
the oscillators in the ensemble have large energy, so that their
angles uniformly sample the possible values. For bound states, by
definition, the energies are low, and hence this limiting form is not
strictly applicable. In particular, the typical amplitudes of the
bound oscillators will be smaller than average, and the true effective
diffusion constant will be less than that of the long time limit. In
addition to its lower value, the diffusion constant will also have
some type of time dependence: In the beginning, when all of the
oscillators in the ensemble have small amplitudes, $D$ will be
correspondingly small. As time increases, even the bound oscillators
will have larger amplitudes and hence the evolution is properly
described by larger values of $D$ (but still smaller than that of the
long time limit).

\subsubsection{Partial Barriers} 

As shown above, the time evolution of the number of bound states
decays exponentially for systems with a strict one-way barrier, i.e.,
for the case in which systems that leave resonance are not allowed to
return.  We now consider the case in which a given system has some
probability $\pback$ of re-entering a bound (resonant) state after
leaving, where $0 \le \pback \le 1$. In this case, all of the systems
are allowed to stay in play, but only a fraction of the systems can
make the transition from an unbound state to a bound state. As shown
below, even when the fraction $\pback \ll 1$, this modification makes
a large qualitative change in the long-term behavior of the system.

We first note that at late times, essentially all of the bound states
in the original problem (without a barrier) are those that have
returned to resonance after leaving. This claim follows directly from
the above results: The fraction of bound states of all types decreases
as $\sim t^{-1/2}$, while the fraction of bound states that have never
left resonance decreases as $\sim \exp[-\gamma t]$, where the decay
rate $\gamma$ depends on the details of the diffusion process. At late
times, however, the power-law behavior wins, and hence most bound
states are due to systems that have returned from higher energy
states.  If the higher energy states are allowed to re-enter resonance
with probability $\pback$, then the fraction of systems in a resonant
state will again follow the $t^{-1/2}$ law, but with a smaller
normalization.  Specifically, the normalization is reduced by the
factor $\pback$ and the long-term solution for the fraction of bound 
states becomes 
\be
\pbound \approx \pback \left( {8 \over \pi D \tau} \right)^{1/2} \, . 
\label{eq:partbarrier} 
\ee 

\begin{figure} 
\figurenum{3} 
{\centerline{\epsscale{0.90} \plotone{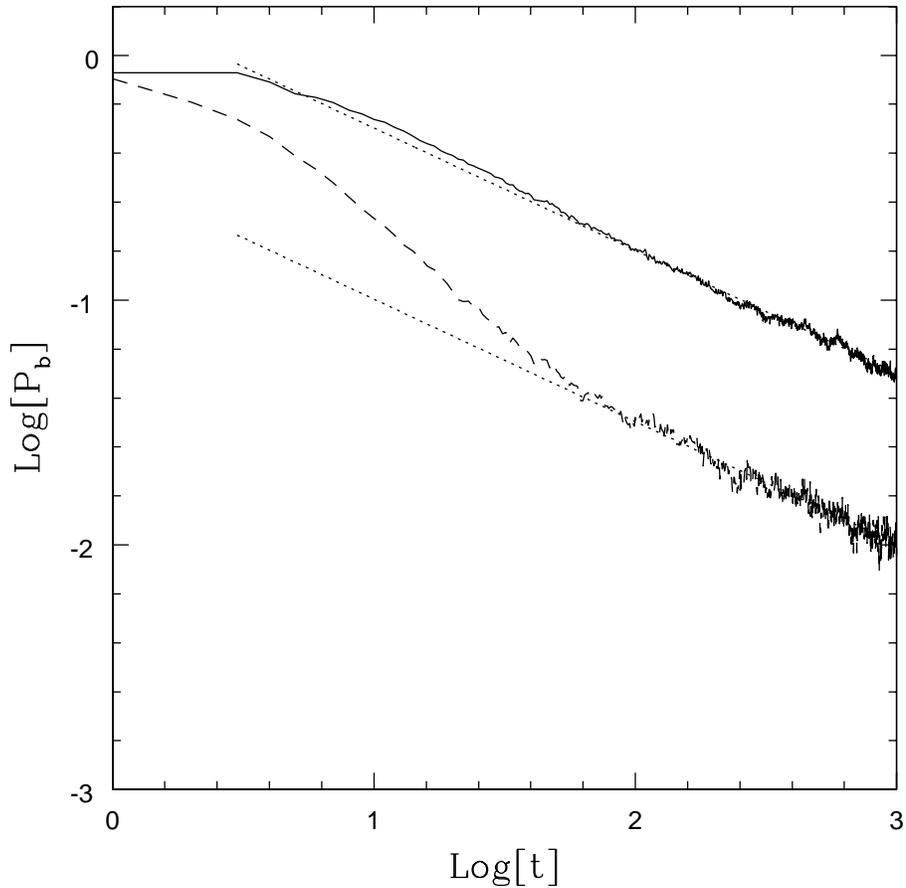} } } 
\figcaption{Time evolution of the fraction of planetary systems that
stay in a mean motion resonance as a function of time. The solid and
dashed curves show the result of $10^4$ realizations of the stochastic
pendulum equation. For the top (solid) curve, the system is allowed to
come in and out of resonance freely; for the bottom (dashed) curve,
the system has only a 20\% chance of returning to a bound state after
leaving.  The two dotted curves shows the expected asymptotic behavior
of the system, i.e., a survival probability of the form $\pbound
\propto t^{-1/2}$. }
\label{fig:partbar}
\end{figure}

This behavior is illustrated in Figure \ref{fig:partbar}, which shows
the time evolution for an ensemble of stochastic pendulums. The upper
two curves show the evolution for the case in which the systems can
freely re-enter a resonant (bound) state after leaving. The bottom two
curves show the fraction of bound states for the case in which the
systems are allowed to have only a 20\% probability of returning to
resonance after leaving.  The solid and dashed curves show the results
of the actual numerical integrations, whereas the dotted curves show
the analytic approximations derived above. Notice the good agreement
between the predicted asymptotic forms and the numerical results. The
fluctuations of the numerically determined fractions about the 
analytic predictions have amplitudes that are consistent with
root-$N$ noise; note that the numerical ensemble contains only
$10^4$ systems and the number of bound states is much smaller by the
end of the time interval shown in Figure \ref{fig:partbar}.

On a related note, one can consider the problem in which systems
leaving a bound state have some probability $p_X$ of never returning
to resonance. Perhaps surprisingly, the long term behavior for this
case is wildly different than for the previous case. Here, for any
nonzero value of $p_X$, the time dependence of the number of bound
states is always an exponential decay.

\subsection{Other Resonances} 

As outlined above, the 2:1 resonance is generally the strongest, and
planets in such a configuration should survive longest in the presence
of turbulence.  For other resonances, the oscillation frequencies are
generally smaller so that the periods of libration are longer. For the
case of the 3:1 resonance, for example, ones finds (MD99) that
\be 
\omega_0^2 \approx - 0.8636 \mu \Omega^2 e^2 \, . 
\label{eq:otheromega} 
\ee 
The negative sign indicates that the stable equilibrium point of the
oscillator occurs at $\varphi = \pi$ rather than $\varphi = 0$.  After
defining $\phi \equiv \varphi - \pi$, the equation of motion for
$\phi$ becomes the same as before.  The above frequency is thus lower
than that of the (simplified) 2:1 resonance by a factor of $\sim
(3/e)^{1/2}$. The potential energy of the oscillator is thus less deep
by a factor $(3/e)^{1/2}$, about 5--6 for typical cases, so that the
system is more easily bounced out of a resonant state by turbulence.

In order to assess the probability of remaining in a bound (resonant)
configuration, one can evaluate the equations derived in \S 2.3 -- 2.5
using the proper frequency given by equation (\ref{eq:otheromega}).
For example, the resulting estimate for the probability of remaining
in resonance has the form of equation (\ref{eq:probtime}), with the
leading coefficient smaller by a factor of $(3/e)^{1/2} \sim 5$. In
other words, at any given time, the probability of a planetary system
remaining in a mean motion resonance, in the face of turbulence, is
inversely proportional to the relevant value of $\omega_0$. For
typical values of the orbital elements of extrasolar planets, where $e
\sim 0.1$, the probability of staying in a 3:1 mean motion resonance
is about 5 times smaller than for a 2:1 resonance.

\subsection{Model Summary} 

The basic analytical model developed here contains three physical
variables that describe the effects of turbulent perturbations on mean
motion resonance. The first variable is the natural oscillation
frequency $\omega_0$ of the resonance (\S 2.1), modeled here as a
pendulum; this quantity represents the oscillation frequency of the
libration angle in the planetary system. The time scale associated
with this frequency is typically $\sim$100 times longer than the
orbital time scale of the planets, and depends on the system
properties in a well-known manner (see \S 2.1 and MD99).  The second
variable is the time scale $\Delta t \approx 8 \pi / \Omega$ (or its
dimensionless counterpart $\Delta \tau \approx 8 \pi \omega_0 /
\Omega$) over which the turbulent perturbations are re-established to
provide an independent realization of the torques. The final variable
is the amplitude of the perturbations. Since the torques and their
corresponding changes in angular momentum can be either positive or
negative, and hence the mean vanishes, the perturbation amplitude can
be characterized by the root mean square $\jrms = \langle (\Delta
J)^2/J^2 \rangle^{1/2}$; this quantity also has a dimensionless
counterpart $\etarms$ given by equations (\ref{eq:etadef}) and
(\ref{eq:etarms}).  Numerical simulations of MHD turbulence indicate
that the dimensionless amplitude is expected to be of order $\etarms
\sim 0.005$ (including the reduction factor due to gap clearing). 

For the simplest version of this model, where systems can freely
random walk back into a resonant state, the three variables
$(\omega_0, \Delta t, \etarms)$ determine the long term evolution of
the ensemble. In particular, the expectation value of pendulum energy
increases linearly with time according to equation (\ref{eq:evst}).
The probability of the system staying (bound) in resonance decreases
as the square root of time according to equation (\ref{eq:probtime}).
Note that both of these results depend only on the dimensionless time
$\tau = \omega_0 t$ and a dimensionless diffusion constant $\diffuse =
\etarms^2 / \Delta \tau$. In other words, the action of the turbulent
fluctuations can be described by a diffusion constant that determines
how the variables of the system -- considered here as a nonlinear
oscillator -- random walk through phase space.

In more realistic versions of the model, where systems cannot easily
return to a resonant state, the long term behavior results in a lower
probability of remaining bound.  As a result, the simple result
described above provides an upper limit on the expected number of
resonant states as a function of time. In the opposite limit in which
systems that leave resonance can never return, the number of bound
states decays exponentially, with varying decay rates, as indicated by
equations (\ref{eq:heursolve}) and (\ref{eq:sepsolve}). For the case
in which systems can re-enter resonance after leaving, but with
reduced probability, the number of bound states decreases as
$t^{-1/2}$ as before, but with reduced normalization, as shown by
equation (\ref{eq:partbarrier}). 

\section{NUMERICAL SIMULATIONS} 

In this section, we consider an ensemble of numerical integrations
motivated by an observed extrasolar planetary system.  For this
exploration, we adopt the outer planets c ($P\sim60\,{\rm d}$) and b
($P\sim30\,{\rm d}$) of the Gliese 876 planetary system (Marcy et al.
2001). These planets are observed to lie deep in the 2:1 mean motion
resonance, with the angles $\theta_1$ and $\theta_2$ librating with
small amplitudes (Laughlin \& Chambers 2001, Rivera et al. 2005).
Indeed, Gliese 876 b and c represent, by far, the most convincing case
of an extrasolar planetary system in mean motion resonance. Among the
25 multiple-planet systems that have been discovered to date, 2:1 mean
motion resonances have also been tentatively identified for HD 82943 b
and c (Gozdziewski \& Maciejewski 2001, Mayor et al. 2004, Lee et al.
2006), HD 128311 b and c (Vogt et al. 2005), and HD 73526 b and c
(Tinney et al. 2005). In each of these three cases, however, the
libration widths of the resonant arguments are very uncertain, and for
HD 128311 and HD 73526, the best dynamical fit to the radial velocity
data shows only a single argument in libration. Following the
discovery of 55 Cancri c and d (Marcy et al. 2002), it was suggested
by several authors (e.g., Ji et al. 2003) that 55 Cancri b and c are
possibly participating in a 3:1 mean motion resonance. Self-consistent
6-body fits to the 55 Cancri data set published by Fischer et al.
(2007), however, show no evidence that 55 Cancri b and c are in 3:1
resonance.\footnote{For an up-to-date $\chi^2$-ranked list of
self-consistent fits to the 55 Cnc data sets see http://oklo.org?p=257
and the links therein.}

In this set of numerical experiments, we start an ensemble of
planetary systems with the observationally determined orbital elements
of GJ876, so that the system begins in a 2:1 mean motion resonance,
and then integrate the system forward in time. The integrations
include additional stochastic fluctuations, which would be present if
the circumstellar disk that formed the planets were still present. We
then monitor how easily the system is knocked out of its resonant
configuration.  The ensemble of numerical integrations reported here
are thus analogous to those done in the previous section with the
stochastic pendulum.  In this case, however, we integrate the full 18
phase space variables (6 orbital elements for each planet and 6 more
for the star) instead of only one variable ($\varphi$) for the
pendulum model. For this system, integrating the motion of the star is
important because its mass is relatively small, only 0.32 $M_\odot$,
whereas the masses of the planets are 0.79 and 2.53 $m_J$, and hence
the system is highly interactive (Laughlin \& Chambers 2001).  As a
result, the GJ876 system is as far from the idealized (one variable)
model of the stochastic pendulum as any planetary system observed to
date; as a result, any agreement found between the simple model and
the full integrations can be considered robust.

The starting orbital elements are taken to be those observed (Rivera
et al. 2005) at the present day; specifically, the epoch for the
initial conditions is JD 2449679.6316, the date for which the first
published Lick Observatory radial velocity was made.  The systems are
then integrated into the future for $2 \times 10^4$ years.  Since the
planets in this particular system have relatively short periods
(60.830 days and 30.455 days for the outer and inner planets,
respectively), the number of orbits is large -- about $2.4 \times
10^5$ orbits of the inner planet.  The integrations are performed with
a Bulirsch-Stoer (B-S) integration scheme (e.g., Press et al. 1992) to
obtain high accuracy with reasonable computational speed. In the
absence of turbulent forcing, the system can be integrated forward in
time for millions of years and is found to stay in resonance, where we
use the following definitions for the libration angles:
\be 
\theta_1 = \lambda_1 - 2 \lambda_2 + 2 (\varpi_1 - \varpi_2) 
\qquad {\rm and} \qquad 
\theta_2 = \lambda_1 - 2 \lambda_2 + (\varpi_1 - \varpi_2) \, ,  
\label{eq:angledef} 
\ee
where the $\lambda_j$ are the mean longitudes and the $\varpi_j$ are
the longitudes of periapse of the two planets.  Note these angles are
defined to be a linear combination of the standard ones (e.g., Lee \&
Peale 2002), and are chosen because they provide a cleaner separation
of the apsidal libration and the mean motions (J. Chambers, private
communication).  We measure the degree of resonance by monitoring the
maximum amplitude of these libration angles over a given interval of
time. For this purpose, we use a fixed monitoring time of 100 years,
which is much longer than the period of the oscillations of the
resonance, about 3.5 years, so that many oscillation periods are
included in the determination of the maximum.  This maximum libration
angle is then tracked as a function of time. For the sake of
definiteness, we consider the systems to be bound (in resonance) when
the maximum angle remains less than $\theta_\ast$ = 90 degrees; when
the maximum angle exceeds $\theta_\ast$ = 90 degrees, we consider the
resonance to be compromised. Note that the exact number of bound
(resonant) states as a function of time depends on the choice of the
angular threshold value $\theta_\ast$; however, the general trends
(shown below) are qualitatively similar for any choice of threshold
angle in the range $90 \le \theta_\ast \le 175$.

The effects of turbulence are introduced by adding small velocity
perturbations at regular time intervals. For the sake of definiteness,
we take the forcing time interval to be 1/3 yr, which is about four
times the period of the inner planet, and is consistent with the time
required for the turbulent torques to be independent (LSA) and with
the model equations of \S 2. In terms of the simple pendulum model of
the system, variations of the stochastic forcing times could be scaled
out of the problem (Appendix A) and would effectively add width to
the distribution of forcing strengths. In these experiments, we take
the size of the velocity perturbations to have the form $\Delta v =
v_A \xi$, where the amplitude $v_A$ = 0.0032 km/s, and where $\xi$ is
a uniformly distributed random variable on the interval $-1 \le \xi
\le 1$.  (This amplitude value was chosen to be a round number, namely
$2 \times 10^{-4}$, in code units, where $G$ = 1, mass is measured in
$M_\odot$, and time is measured in years.)  Both the $\hat x$ and
$\hat y$ components of the velocity are perturbed (independently) in
this manner, but the vertical velocity component is left unchanged.
Including both components of the velocity perturbation, we find that
the fractional change in speed, and hence angular momentum, has
amplitude $\jrms \sim 10^{-4}$, consistent with the benchmark 
value of equation (\ref{eq:bench}).

\begin{figure} 
\figurenum{4} 
{\centerline{\epsscale{0.90} \plotone{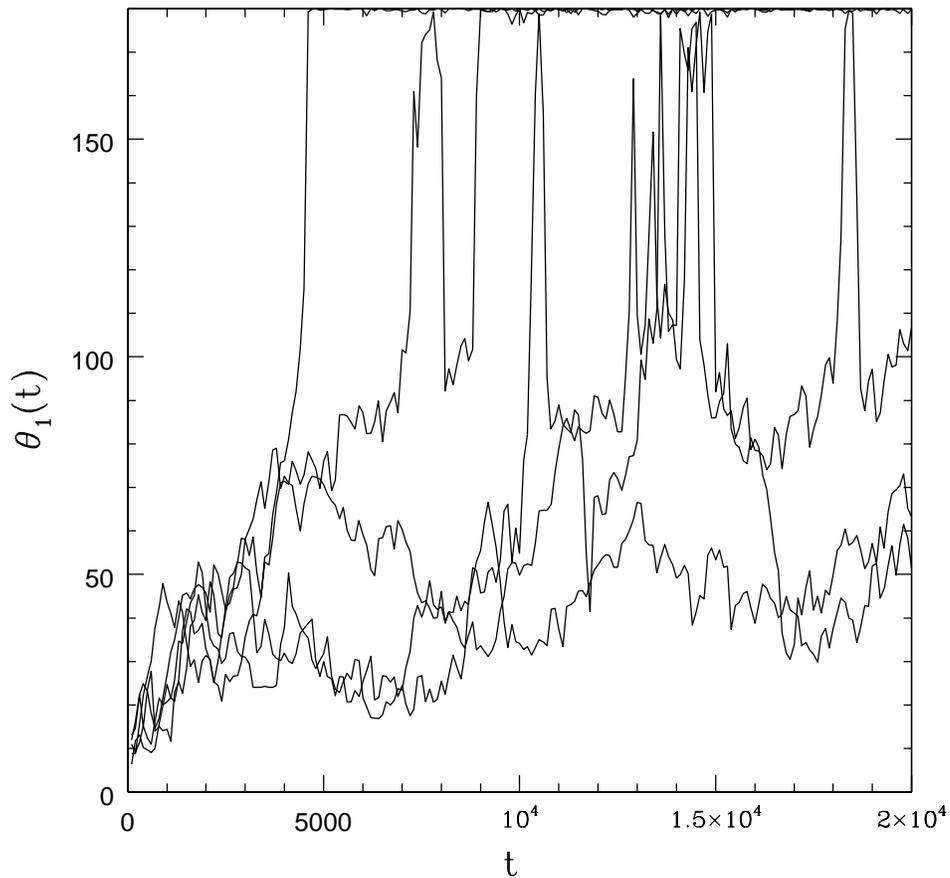} } } 
\figcaption{Time evolution of the resonance angles for a collection of
planetary systems based on the observed system GJ876. The systems are
subjected to turbulent forcing as described in the text. The five
curves show the first resonance angle $\theta_1$ as a function of time
(given here in years) for the five different realizations of the 
turbulent fluctuations. }
\label{fig:phievolve} 
\end{figure} 

\begin{figure}
\figurenum{5}  
{\centerline{\epsscale{0.90} \plotone{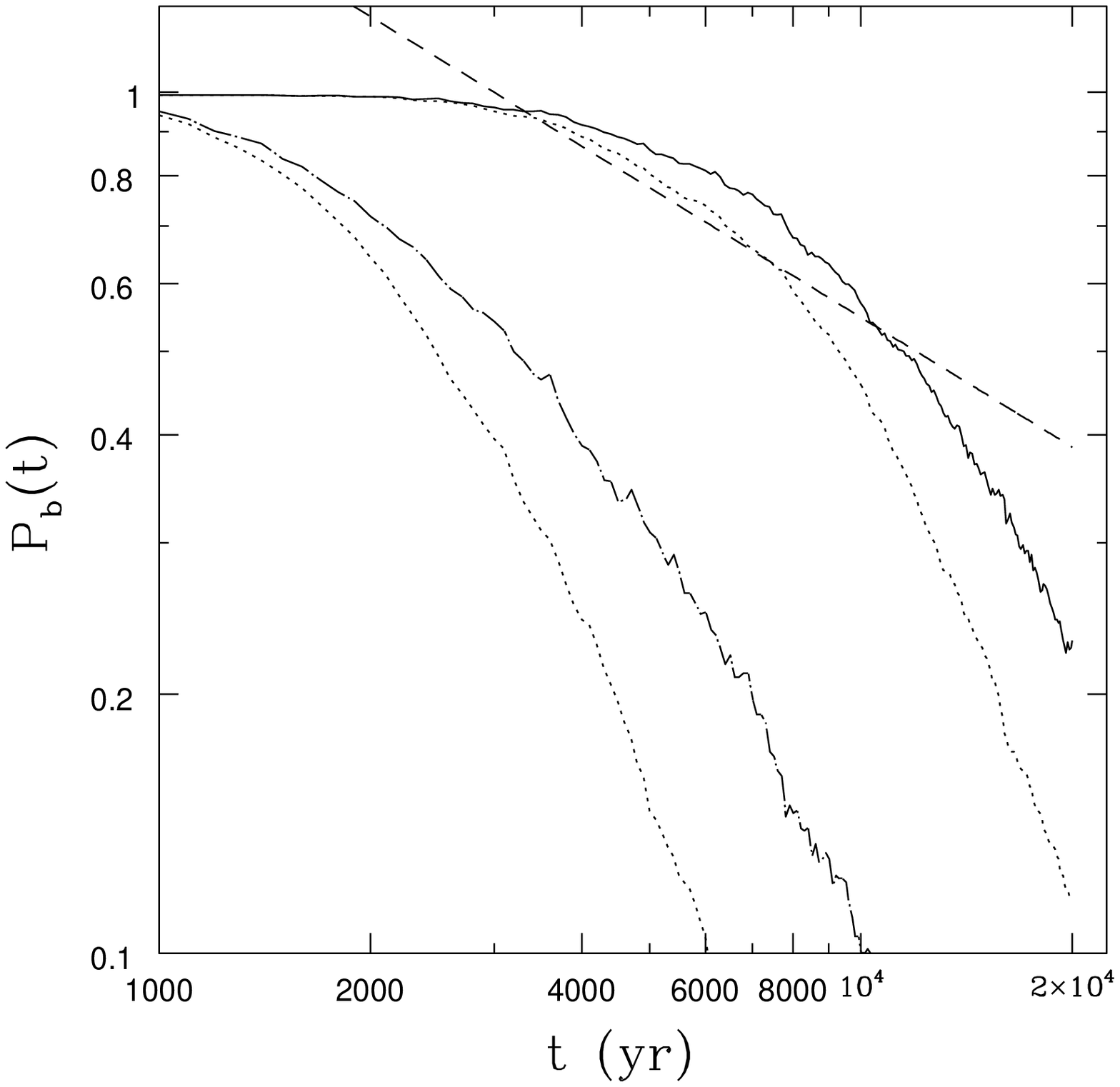} } } 
\figcaption{Time evolution of the fraction of bound states, as
measured by the maximum excursions of the resonance angles, for a
collection of planetary systems based on the observed system GJ876. 
The systems are subjected to turbulent forcing as described in the
text. The solid and dot-dashed curves show the fraction of systems
that remain in resonance as a function of time (in years) for the
first and second resonance angles, respectively.  The dotted curve
shows the fraction of bound states for the first resonance angle that
would result if systems that leave resonance could never return to a
bound state. For reference, the dashed curve shows the power-law
behavior $P_{\rm b} \propto t^{-1/2}$ expected for a simple stochastic
pendulum. Note that the survival probability $P_{\rm b} \ll 0.01$ if 
we extrapolate this result out to time scales of $\sim$1 Myr. } 
\label{fig:gj876} 
\end{figure} 

The results from one ensemble of simulations are shown in Figures
\ref{fig:phievolve} and \ref{fig:gj876}.  Figure \ref{fig:phievolve}
shows the time evolution of the maximum libration angle (as defined
above) for five different trials, i.e., five different realizations of
the turbulent fluctuations. Note that the systems can re-enter bound
states -- defined here to be maximum libration angles less then
$\theta_\ast$ = 90 degrees -- after leaving.  Nonetheless, the results
show a clear trend for systems to preferentially leave resonance,
rather than return, so the number of bound states decreases rapidly
with time.

We have performed an ensemble of 900 integrations of the system, 
where each numerical experiment uses a different realization of the 
stochastic perturbations due to turbulence (with this sample size,
root-$N$ fluctuations are $\sim$ 1/30 $\sim$ 0.03).  Figure
\ref{fig:gj876} shows the fraction of bound states as a function of
time.  The solid and dot-dashed curves show the fraction of systems
that remain in resonance as a function of time (in years) for the
first and second resonance angles, respectively.  The dotted curves
shows the corresponding fraction of bound states that would remain if
the systems that leave resonance could never return to a bound
state. These results clearly indicate that the actual fraction of
bound states is systematically larger than the fraction of bound
states that would result if systems were never allowed to return to a
resonant state after leaving (compare the dotted and solid curves in
Figure \ref{fig:gj876}). In other words, planetary systems can ---
sometimes --- random walk back into a resonant configuration after
leaving.  This behavior is consistent with that found for the
evolution of individual systems (shown in Fig. \ref{fig:phievolve}).
Finally, the dashed curve shows the power-law behavior $\pbound
\propto t^{-1/2}$ for reference.  Note that the time evolution of the
fraction of bound states falls well below the prediction of a pure
power-law with $\pbound \sim t^{-1/2}$. The long term behavior of the
number of bound states seems to follow a steeper power-law, but does
reach a full exponential decay during the time window studied (which
would be expected if systems never return to resonance).

As an order of magnitude check on the time scales, note that the
fractional change in velocity per stochastic kick has amplitude
$\Delta v \sim 10^{-4} v_{orb}$, where $v_{orb}$ is the orbital speed
of the inner planet.  As outlined in \S 2, however, the energy
associated with the potential well of the resonance is much smaller
than that of orbit.  If we denote the change in velocity required to
leave resonance as $v_{res}$, then $v_{res} \sim 0.01 v_{orb}$, which
in turn implies $\Delta v \sim 0.01 v_{res}$. If $\Delta v$
accumulates as a random walk, the system thus requires $\sim10^4$
stochastic forcing kicks to account for enough energy to compromise
resonance, and this number translates into a time scale of $\sim3000$
yr. Indeed, individual systems (see Fig. \ref{fig:phievolve}) and the
ensemble (see Fig. \ref{fig:gj876}) evolve on a comparable time scale.

\section{DISCUSSION AND CONCLUSION} 

The main finding of this paper is that mean motion resonance in 
planetary systems is readily compromised through the action of
turbulent fluctuations in circumstellar disks.  If MRI operates and
the accompanying turbulent torques are common during the epoch of
planetary migration, then planetary systems in mean motion resonances
are predicted to be rare. Specifically, even for the most favorable
case in which the system can freely random walk in and out of
resonance, only a small percentage of the solar systems that produce
pairs of planets in a resonant configuration will remain in resonance
over typical disk lifetimes of $\sim1$ Myr if turbulence is present.
On the other hand, if planetary systems in mean motion resonance are
found to be common, then this analysis implies that turbulence, and
hence MRI, would have a severely limited duty cycle.

We have addressed this problem using both direct numerical
integrations of sample solar systems and through the construction of
model equations that represent an ensemble of stochastically driven
pendulums. This latter approach allows for the derivation of a number
of analytic results that elucidate the basic physics of the problem
--- solar systems being driven from their resonant states by turbulent
forcing.  For this class of systems, the energy of the ensemble of
stochastic oscillators increases linearly with time (see Fig.
\ref{fig:enertime}, eq. [\ref{eq:evst}], and Appendix B), the
distribution of energy for the ensemble spreads according to the
solution of equation (\ref{eq:edistrib}), and the probability of a
planetary system remaining in resonance decreases as $t^{-1/2}$ (see
Fig.  \ref{fig:pdecay}). This law holds both for the case in which
systems can freely re-enter resonance, and for the case in which they
re-enter resonance with a given (fixed) probability, although the
normalization is lower for the latter case (eq. [\ref{eq:partbarrier}]). 
If systems that leave resonance can never return to a bound state,
then the time evolution of the fraction of bound states is a decaying
exponential (\S 2.5).

The results from our numerical integrations indicate that the fully
interactive system (with 18 phase space variables) behaves in a
qualitatively similar manner to the stochastic pendulum (with one
variable). In particular, the number of bound states (systems in
resonance) decreases with time, where the evolution of the ensemble of
systems is diffusive, and where the diffusion constant depends on the
square of the fluctuation amplitude and inversely on the time scale of
the turbulent forcing.  The resulting function that describes the
decrease in the number of resonant states is intermediate between the
$t^{-1/2}$ power-law behavior of the simplest pendulum system and the
decaying exponential form that applies for systems that can never
re-enter resonance after becoming unbound. The numerical simulations
show that systems can indeed random walk back into a resonant state
after leaving, consistent with the finding that the number of bound
states does not decay exponentially. On the other hand, the return to
resonance is relatively rare, in particular more unlikely than for the
(one variable) stochastic pendulum, a result that is consistent with
the fraction of bound states falling below the $t^{-1/2}$ law. This
latter power-law behavior --- the solution to the the long term
behavior of a stochastic pendulum -- thus provides an upper limit on
the probability of resonant states as a function of time.

The most important astronomical implication of this work is a
quantitative estimate of the rarity of solar systems surviving in a
resonant configuration.  Our results can be summarized by the
following expression, which represents our working estimate for the
fraction of surviving bound (resonant) states:
\be
\pbound \approx {\pback \over \etarms} 
\left( {32 \over \pi \norb} \right)^{1/2} \, 
\equiv {C \over \norb^{1/2} } \, , 
\label{eq:netresult} 
\ee
where $\pback$ is the probability of returning to resonance, $\etarms$
is set by the amplitude of the fluctuations, and $\norb$ is the number
of orbits over which the turbulence is active.  In the second equality
we have defined the dimensionless quantity $C = (32/\pi)^{1/2} \pback
/ \etarms$.  For typical turbulent amplitudes, $\etarms \approx 0.005$,
and hence $C \approx 640 \pback$.  Given that returning to resonance
requires somewhat special conditions, we expect $\pback$ to be small, 
i.e., $\pback \ll 1$. This expectation is borne out in our numerical 
simulations (\S 3), which show that $\pbound$ decreases rapidly with
time. As a result, we expect that the constant $C$ will be of order
$\sim 10$ (perhaps as large as $\sim50$) so that $\pbound \approx 10 
/ \norb^{1/2}$. If turbulence is present for a substantial fraction
of the disk lifetime, then $\norb = 10^6 - 10^8$ and $\pbound$ = 0.001
-- 0.01, i.e., mean motion resonances will be rare.

In addition to the prediction that resonant states are rare, this work
has other astronomical implications: For the particular case of the
Gliese 876 system, if no eccentricity damping is assumed for the epoch
of planet formation and migration, then the current eccentricities and
libration widths of the system suggest that the system migrated only a
small distance (8\% of the initial semi-major axes) while in resonance
(Lee \& Peale 2002).  This result is consistent with any turbulence in
the disk of the GJ876 system having had little time or ability to
provide substantial perturbations on the resonant arguments.  On the
other hand, the observed HD 128311, HD 82943 and HD73526 planetary
systems are all consistent with having received significant exposure
to turbulent perturbations, but not enough to have destroyed their
librations completely. Once a large sample of multiple-planet systems
has been assembled, we should gain further insight from the
distribution of observed resonant widths, in addition to the observed
fraction of planets in mean motion resonance. In particular, if
planets can random walk back into the resonance after leaving, as
indicated by both our numerical integrations and model equations, the
observational sample should show a definite preference for systems
with large libration widths. This trend is exactly what we observe for
the tenuous resonances in the three systems other than Gliese 876.

This analysis also outlines the basic physical mechanism by which
turbulence affects mean motion resonance. We have modeled the process
through a pendulum equation (which represents the resonance) with the
addition of stochastic forcing (which represents the turbulence).
This formulation applies specifically to the class of systems for
which the turbulent fluctuations provide forcing kicks that are fully
independent, i.e., where both the amplitude of the fluctuations and
the time intervals are not correlated. In this case, the results are
largely independent of the distributions, and depend primarily on the
expectation values of the amplitudes and time intervals of the
turbulent forcing. These latter quantities set the value of the
effective diffusion constant (for an analogous problem, see Fatuzzo \&
Adams 2002, which considers ambipolar diffusion with turbulent
fluctuations in molecular cloud cores).

As an intuitive check on the results presented herein, consider the
following heuristic argument. In a stochastic pendulum, the system
accumulates angular momentum as a random walk, so the change in
angular momentum after $N_{\rm step}$ steps is given by $N_{\rm step}
(\Delta J)_k^2$, where $(\Delta J)_k$ is the typical amplitude of the
angular momentum fluctuations. In order of magnitude, the amount of
angular momentum in the resonant system (when bound) is given by
$(J_{\rm orb} \omega_0 / \Omega)^2$.  The combination of these two
expressions indicates that the criterion for compromising resonance
can be written in the form $N_{\rm step} \ge 
[(\Delta J)_k/J_{\rm orb}]^{-2} (\omega_0/\Omega)^2$.  In order of
magnitude, $(\Delta J)_k/J_{\rm orb} \sim 10^{-4}$ and $\omega_0/\Omega 
\sim 10^{-2}$, and hence the required number of steps is of order
$N_{\rm step} \sim 10^4$.  Indeed, when the number of stochastic steps
(measured here by the number of orbits $\norb$) exceeds this limit,
the probability of remaining in a bound state decreases (see
eqs. [\ref{eq:probtime}] and [\ref{eq:netresult}]).

Notice also that the effective diffusion constant $D \propto (\Delta
J)_k^2 / (\Delta t)_k$.  In order to set the proper value of the
constant $D$, one must choose the time interval to be long enough so
that the fluctuations are independent and short enough that the
changes in angular momentum $(\Delta J)$ increase linearly with time
(where we now suppress the index $k$). Suppose, for example, one were
to choose a longer time interval to sample the fluctuations, say
$(\Delta t) \gg (\Delta t)_0$, where $(\Delta t)_0$ is the time
interval for optimal independent sampling. The change in angular
momentum over one sampling interval would then increase as a random
walk (even within the time interval), i.e., $(\Delta J) = (\Delta J)_0
[(\Delta t) / (\Delta t)_0]^{1/2}$.  The effective value of the
diffusion constant then scales according to $D \propto (\Delta J)^2 /
(\Delta t) = (\Delta J)_0^2 [ (\Delta t) / (\Delta t)_0] / (\Delta t)
= (\Delta J)_0^2 / (\Delta t)_0$. We thus recover the correct value,
provided that the changes in angular momentum are computed properly.
In this paper, we have taken the time interval $\Delta t$ to be four
times the orbital period of the inner planet (twice the period of the
outer planet), consistent with the calculations of LSA (see also
Nelson 2005).  The above considerations imply that the diffusion
constant $D$ is inversely proportional to $\Delta t$, and the fraction
of bound states scales as $D^{-1/2}$, so that our quoted results are
only weakly dependent on any uncertainty in the specification of
$\Delta t$.

Although our general conclusion is robust -- namely, that turbulence
easily compromises mean motion resonances -- a good deal of additional
work should be done. The parameter space available for these planetary
systems is enormous and further exploration should be carried out for
different solar system architectures. In particular, the degree to
which the full numerical integration agree with the simple model
equations should be considered, including a determination of the solar
system properties required for consistency. This treatment assumes
that the planets have already formed, and have already migrated to
their (nearly) final radial locations, and then considers the effects
of turbulence on mean motion resonances. In actuality, planets can be
subjected to turbulent fluctuations as they form, and as they migrate
inward and become locked into resonance.  These processes of planet
formation, migration, resonance locking, and turbulent forcing should
thus be studied concurrently. In addition, the differences between the
results of our full numerical integrations (which include all 18 phase
space variables of the system) and those of the idealized stochastic
pendulum (which has only one degree of freedom) poses a mathematically
interesting issue for further study.

\acknowledgements
This work was initiated during a sabbatical visit of FCA to U. C.
Santa Cruz; we would like to thank both the Physics Department at the
University of Michigan and the Astronomy Department at the University
of California, Santa Cruz, for helping to make this collaboration
possible. We thank REU student Jeff Druce for performing supporting
calculations and thank an anonymous referee for many insightful
comments that improved and clarified the paper.  This work was
supported in part by the Michigan Center for Theoretical Physics. FCA
is supported by NASA through the Origins of Solar Systems Program via
grant NNX07AP17G.  AMB is supported by the NSF through grants
CMS-0408542 and DMS-604307.  GL is supported by the NSF through CAREER
Grant AST-0449986, and by the NASA Planetary Geology and Geophysics
Program through grant NNG04GK19G.

\appendix
\section{RESCALING FOR VARIABLE TIME INTERVALS} 

In this Appendix, we show that if the stochastic forcing impulses
occur at irregular time intervals, then the time variable can be
rescaled to make the stochastic forcing kicks occur at regular
(periodic) intervals. As a compensating result, however, the
distribution of the forcing strengths becomes wider and the natural
oscillation frequency of the system takes on a distribution of values.

We begin by writing the stochastic pendulum equation in the form 
\be
{d^2 \varphi \over d t^2} + \left[ \omega_0^2 + q_k Q(\lambda_k t) 
\right] \sin \varphi = 0 \, , 
\ee
where $Q$ is a periodic function and $q_k$ determines the forcing 
strength for the $kth$ time interval; the variable $\lambda_k$ 
has unit mean, but allows for different time intervals for the 
stochastic forcing terms. Notice that we generally use periodic 
Dirac-delta functions to specify the forcing time, but this 
argument can be generalized to include any periodic function 
$Q$. If we re-scale the time variable for each cycle according to 
\be
t \to \lambda_k t \, , 
\ee
then the stochastic pendulum equation takes the form 
\be
\lambda_k^2 {d^2 \varphi \over d t^2} + \left[ \omega_0^2 + q_k Q(t) 
\right] \sin \varphi = 0 \, , 
\ee 
where the time variable appearing in this equation is the rescaled 
one. If we then re-scale the remaining parameters $(\omega_0, q_k)$ 
according to the relations 
\be
\omega_0 \to \omega_0 / \lambda_k \equiv \omega_k 
\qquad {\rm and} \qquad q_k \to q_k/\lambda_k^2 \equiv {\tilde q}_k \, , 
\ee 
the stochastic pendulum equation has the same form as that for the
case of perfectly periodic forcing terms. This result is thus the
analog of Theorem 1 of Adams \& Bloch (2008) for the stochastic Hill's
equation. Notice that in this case the forcing strength $q_k$ acquires
a new (generally wider) distribution to accommodate the variation in
forcing periods, and, the natural oscillation frequency $\omega_0$
becomes a random variable.

\section{ALTERNATE DERIVATION OF THE LINEAR TIME DEPENDENCE FOR THE 
MEAN ENERGY}

In this Appendix, we present an alternate derivation of the result
that the energy of a stochastically driven pendulum tends to increase
linearly with time (in the limit of long times). This result is thus 
consistent with equation (\ref{eq:evst}) in the text. 

We begin by considering the pendulum equation with a periodic
delta-function forcing term, as described in the text (\S 2).  The
angle $\varphi$ itself must be continuous at the moment of forcing,
but the derivative of $\varphi$ obeys a jump condition that can be
written in the form
\be 
{d \varphi \over d\tau}\Bigg|_+ = 
{d \varphi \over d\tau}\Bigg|_- - q_k \sin \varphi_k \, , 
\ee
where we have denoted the value of the angle at the $kth$ cycle
boundary as $\varphi_k$. Since the angle $\varphi$ must be a
continuous function across the boundary, the potential energy is also
a continuous function, and hence the total energy experiences discrete
jumps of the form
\be
\Delta E = - q_k \sin \varphi_k {d \varphi \over d\tau}\Bigg|_- 
+ {1 \over 2} q_k^2 \sin^2 \varphi_k \, . 
\ee
Over long times, the first term in the above equation averages to
zero, whereas the second term averages to $\langle q_k^2 \rangle / 4$. 
The mean energy thus grows as
\be
E \approx {\langle q_k^2 \rangle \over 4 \Delta \tau } \tau \, , 
\ee
where $\Delta t$ is the time interval of the stochastic forcing.  Not
only does this argument reproduce the linear growth of the mean energy, 
as found using the Fokker-Planck analysis in the text, but it also 
results in the same coefficient if we identify the effective diffusion 
constant $D = \langle q_k^2 \rangle / (\Delta \tau)$.

\end{document}